\documentclass[useAMS,usenatbib,usegraphicx]{mn2e}

\usepackage{psfig}

\def\aj{AJ}%
\def\apj{ApJ}%
\def\apjl{ApJ}%
\def\aap{A\&A}%
\def\aaps{A\&AS}%
\def\mnras{MNRAS}%
\def\pasj{PASJ}%
\def\nat{Nature}%
\title[Synchrotron emission from the blazar PG~1553+113.]
{Synchrotron emission from the blazar PG~1553+113. An analysis of its flux and polarization variability.
}
\author[C. M. Raiteri et al.] 
{C.~M.~Raiteri              $^{ 1}$\thanks{E-mail:raiteri@oato.inaf.it},
F.~Nicastro                 $^{ 2}$,
A.~Stamerra                 $^{ 1,3,4}$,
M.~Villata                  $^{ 1}$,
V.~M.~Larionov              $^{ 5,6}$,
\newauthor
D.~Blinov                   $^{ 7,8,5}$,
J.~A.~Acosta-Pulido         $^{ 9,10}$,
M.~J.~Ar\'evalo             $^{ 9,10}$,
A.~A.~Arkharov              $^{ 6}$,
\newauthor
R.~Bachev                   $^{11}$,
G.~A.~Borman                $^{12}$,
M.~I.~Carnerero             $^{ 1}$,
D.~Carosati                 $^{13,14}$,
M.~Cecconi                  $^{14}$,
\newauthor
W.-P.~Chen                  $^{15}$,
G.~Damljanovic              $^{16}$,
A.~Di~Paola                 $^{ 2}$,
Sh.~A.~Ehgamberdiev         $^{17}$,
\newauthor
A.~Frasca                   $^{18}$,
M.~Giroletti                $^{19}$,
P.~A.~Gonz\'alez-Morales    $^{ 9,10}$,
A.~B.~Gri\~non-Mar\'in      $^{ 9,10}$,
\newauthor
T.~S.~Grishina              $^{ 5}$,
P.-C.~Huang                 $^{15}$,
S.~Ibryamov                 $^{11,20}$,
S.~A.~Klimanov              $^{ 6}$,
\newauthor
E.~N.~Kopatskaya            $^{ 5}$,
O.~M.~Kurtanidze            $^{21,22}$,
S.~O.~Kurtanidze            $^{21}$,
A.~L\"ahteenm\"aki          $^{23,24}$,
\newauthor
E.~G.~Larionova             $^{ 5}$,
L.~V.~Larionova             $^{ 5}$,
C.~L\'azaro                 $^{ 9,10}$,
G.~Leto                     $^{18}$,
I.~Liodakis                 $^{ 7,8}$,
\newauthor
C.~Mart\'inez-Lombilla      $^{ 9,10}$,
B.~Mihov                    $^{11}$,
D.~O.~Mirzaqulov            $^{17}$,
A.~A.~Mokrushina            $^{ 6}$,
\newauthor
J.~W.~Moody                 $^{25}$,
D.~A.~Morozova              $^{ 5}$,
S.~V.~Nazarov               $^{12}$,
M.~G.~Nikolashvili          $^{21}$,
\newauthor
J.~M.~Ohlert                $^{26,27}$,
G.~V.~Panopoulou            $^{ 7,8}$,
A.~Pastor~Yabar             $^{ 9,10}$,
F.~Pinna                    $^{ 9,10}$,
\newauthor
C.~Protasio                 $^{ 9,10}$,
N.~Rizzi                    $^{28}$,
A.~C.~Sadun                 $^{29}$,
S.~S.~Savchenko             $^{ 5}$,
E.~Semkov                   $^{11}$,
\newauthor
L.~A.~Sigua                 $^{21}$,
L.~Slavcheva-Mihova         $^{11}$,
A.~Strigachev               $^{11}$,
M.~Tornikoski               $^{23}$,
\newauthor
Yu.~V.~Troitskaya           $^{ 5}$,
I.~S.~Troitsky              $^{ 5}$,
A.~A.~Vasilyev              $^{ 5}$,
R.~J.~C.~Vera               $^{23,24}$,
O.~Vince                    $^{16}$,
\newauthor
and R.~Zanmar~Sanchez        $^{18}$
}
\begin{document}
\pagerange{\pageref{firstpage}--\pageref{lastpage}} \pubyear{2015}
\maketitle
\label{firstpage}
\begin{abstract}
In 2015 July 29 -- September 1 the satellite {\em XMM-Newton} pointed at the BL Lac object PG~1553+133 six times, collecting data for 218 hours. 
During one of these epochs, simultaneous observations by the {\em Swift} satellite were requested to compare the results of the X-ray and optical--UV instruments. 
Optical, near-infrared and radio monitoring was carried out by the Whole Earth Blazar Telescope (WEBT) collaboration for the whole observing season. 
We here present the results of the analysis of all these data, together with an investigation of the source photometric and polarimetric behaviour over the last three years. 
The 2015 EPIC spectra show slight curvature and the corresponding light curves display fast X-ray variability with a time scale of the order of 1 hour. 
In contrast to previous results, during the brightest X-ray states detected in 2015 the simple log-parabolic model that best-fits the {\em XMM-Newton} data also reproduces reasonably well the whole synchrotron bump, suggesting a peak in the near-UV band. 
We found evidence of a wide rotation of the polarization angle in 2014, when the polarization degree was variable, but the flux remained almost constant. This is difficult to interpret with deterministic jet emission models, while it can be easily reproduced by assuming some turbulence of the magnetic field.
\end{abstract}

%

\section{Introduction}

The active galactic nuclei known as ``blazars" include flat-spectrum radio quasars and BL Lac objects. They show strong flux variability over all the electromagnetic spectrum, high and variable polarization, superluminal motion of the radio knots, brightness temperatures exceeding the Compton limit. These properties are explained by assuming that blazar emission comes from a plasma jet oriented close to the line of sight, implying relativistic beaming of the radiation \citep{bla79}. This is produced by synchrotron and likely inverse-Compton processes at low and high energies, respectively \citep{koe81}.

The source PG 1553+113 is a high-energy peaked BL Lac object (HBL), which means that its spectral energy distribution (SED) shows a synchrotron bump extending from the radio to the X-rays, and an inverse-Compton bump from X-rays to TeV energies.
For these sources, the inverse-Compton radiation is usually assumed to originate from the scattering of synchrotron photons by the same relativistic electrons that produce the synchrotron emission \citep[SSC models; e.g.][]{mas97}.
The redshift of PG 1553+113 has been investigated in several ways, but has not been firmly established yet. Constraints derived from the analysis of the intergalactic absorption features set it in the range $0.43 < z < 0.58$ \citep{dan10}.
Lower limits to the redshift were also obtained from spectroscopy, $z > 0.12$ \citep{sha13} and $z > 0.3$ \citep{lan14}.
Studies of the very-high-energy (VHE, $E>100 \, \rm GeV$) spectrum suggested $z<0.5$ \citep{orr11}, $z = 0.49 \pm 0.04$ \citep{abr15}, or $z \sim 0.4$ \citep{ale15}.
The source has recently gained particular attention because of the $\sim 2$ year quasi-periodicity of its flares \citep{ack15} and because of its possible association with neutrino signals detected by IceCUBE \citep[e.g.][]{pad16,rig16}.

The SED modelling has produced different results.
\citet{abd10_1553} obtained satisfactory fits to the broad-band source SED, from the radio to the VHE band, with a one-zone SSC model where the energy distribution of the electrons producing the synchrotron radiation is modeled by a three-component power law. 
Different source states were accounted for by only changing the parameters determining this law, while \citet{ale12} had to fine-tune six parameters
to explain the noticeable SED variations.
A one-zone SSC model was also applied to the broad-band SED observed during the 2012 flaring state by \citet{ale15}.

In 2013 the Whole Earth Blazar Telescope\footnote{http://www.oato.inaf.it/blazars/webt/} (WEBT) coordinated a multiwavelength campaign on PG~1553+113 in support of MAGIC observations. The results of the WEBT radio-to-optical monitoring were presented in \citet{rai15} and used together with observations by {\em Swift} and {\em XMM-Newton} to analyse the source synchrotron emission. The picture that emerged was enigmatic, because the connection between the UV and X-ray spectra seemed to require a double curvature. Possible interpretations were discussed, favouring variations of the viewing angle of the jet emitting regions, whose effects were tested with an inhomogeneous SSC helical jet model \citep{vil99,rai09}. Different SED shapes corresponding to various brightness states were matched by changing only three parameters defining the jet geometry.

In 2015, nearly 218 hours of {\em XMM-Newton} observing time on PG~1553+113 were awarded to investigate the warm-hot intergalactic medium \citep[WHIM; e.g.][and references therein]{nic13}.
At the same time, this offered an extraordinary opportunity to study in detail the emission of the blazar itself and, in particular, the problem of the spectral connection between the UV and X-ray regions discussed by \citet{rai15}.
To maximize the scientific output, the WEBT organized a new multifrequency campaign involving optical, near-infrared, and radio observing facilities. Moreover, {\em Swift} target of opportunity observations simultaneous to one of the {\em XMM}-Newton pointings were activated to check for cross-calibration between the optical-UV and X-ray detectors. 
MAGIC observations were also scheduled in the first half of August. Their results will be reported elsewhere.

In this paper we present the results of the six {\em XMM-Newton} guest-observer (GO) pointings at PG 1553+113 in 2015 (Section 2), together with those of the {\em Swift} observations (Section 3) and WEBT continuous monitoring (Section 4) in the same period. We discuss the SEDs built with contemporaneous data in the six {\em XMM-Newton} epochs (Section 5). Moreover, we analyse the long-term (2013--2015) optical photometric and polarimetric behaviour of the source in conjunction with that observed in the X-ray band (Section 6) and discuss a possible interpretation (Section 7).
Summary and conclusions are outlined at the end (Section 8).

\section{Observations by {\em XMM-Newton}}
Six long GO observations of PG~1553+113 were performed by XMM-Newton in 2015, between July 29 and September 1 (PI: F.~Nicastro).
Their logbook is given in Table \ref{xmm_log}.
The total observing time was 784100 s.

\begin{table}
\caption{Logbook of the {\em XMM-Newton} GO observations of PG~1553+113 in 2015 (PI: F.~Nicastro).}
\label{xmm_log}
\begin{tabular}{lllr}
\hline
Rev. & Start & End & Duration (s) \\
\hline
2864 &	2015-07-29 19:57:33 &	2015-07-31 10:24:13 & 138400 \\	         
2866 &	2015-08-02 19:40:00 &	2015-08-04 10:15:00 & 138900 \\	      
2867 &	2015-08-04 19:32:00 &	2015-08-06 10:07:00 & 138900 \\	       
2869 &	2015-08-08 19:12:07 &	2015-08-10 09:47:07 & 138900 \\	          
2873 &	2015-08-16 18:52:06 &	2015-08-17 19:52:06 &  90000 \\	      
2880 &	2015-08-30 17:52:29 &	2015-09-01 08:29:09 & 139000 \\	
\hline
\end{tabular}
\end{table}

\subsection{EPIC}
The European Photon Imaging Camera (EPIC) onboard {\em XMM-Newton} carries three detectors: MOS1, MOS2 \citep{tur01} and pn \citep{str01}.
During the GO observations of PG 1553+113 in 2015 the pn detector was set in full frame mode\footnote{With the only exception of the observation on August 16--17, when it was set to small window.} with thin filter; MOS1 operated in small window mode with thick filter, and MOS2 in small window mode with thin filter.
We ran the {\tt emproc} and {\tt epproc} tasks of the {\sevensize \bf SAS} package, version 14.0.0, to reduce the data. 

\subsubsection{Spectra}
\label{epic}
Good time intervals with low background counts were selected by asking that the count rate of high-energy events ($> 10 \, \rm keV$) was less than 0.35 and 0.40 counts $\rm s^{-1}$ on the MOS and pn detectors, respectively. 
We extracted source counts from a circular region with 25--35 arcsec radius and the background counts from a source-free circle with 50--70 arcsec radius.
The {\tt epatplot} task was used to verify the presence of pile-up. This was not a problem for the MOS detectors, but in the case of the pn data we had to pierce the source counts extraction region, excluding a central circle with 10 arcsec radius\footnote{This was not necessary for the August 16--17 data.}. 
We selected only single and double events (PATTERN$<$=4), which are the best calibrated ones, and rejected events next to either the edges of the CCDs or bad pixels (FLAG==0). We produced effective area files with the task {\tt arfgen} and redistribution matrices files with {\tt rmfgen}. 
We used the task {\tt grppha} of the {\sevensize \bf Heasoft} package to bin each spectrum and associate it with the corresponding background, redistribution matrix (rmf), and ancillary (arf) files. We set a binning of at least 25 counts for each spectral channel to use the $\chi^2$ statistic.
All spectra were analysed with {\sevensize \bf  Xspec} version 12.9.0 between 0.3 keV and 10 keV in the case of MOS, and between 0.3 keV and 12 keV in the pn case. We fitted the three EPIC spectra of each observation first separately and then together. In the latter case we allowed for some offset between detectors and normalized with respect to MOS1. 

The spectra were fitted with both an absorbed power law $N(E)=N_0 \, E^{-\Gamma}$, where $N_0$ represents the number of photons $\rm keV^{-1} \, cm^{-2} \, s^{-1}$ at 1 keV, and an absorbed log-parabolic model $$N(E)=N_0 \, (E/E_{\rm s})^{[-\alpha-\beta \log (E/E_{\rm s})]},$$
where $E_{\rm s}$ is a scale parameter that we fixed equal to 1 keV. 
In all cases we adopted the \citet{wil00} elemental abundances.
An absorbed power law with $N_{\rm H}$ fixed at the Galactic value of $3.72 \times 10^{20} \rm \, cm^{-2}$ \citep[from the LAB survey;][]{kal05} gave statistically bad fits, with $\chi^2_{\rm red}$ of 1.25, 1.29, 1.21, 1.07, 1.27, and 1.49 for the six epochs. We obtained good fits with a power-law with free absorption and acceptable fits with a log-parabola with absorption fixed at the Galactic value.
The results of the spectral fitting with these two models are shown in Tables \ref{epic_pow} and \ref{epic_logpar}.

\begin{table*}
\caption{Results of the EPIC spectral fitting with a power-law model with free absorption}
\label{epic_pow}
\begin{tabular}{lllll}
\hline
Date &  $N_{\rm H} \,  (10^{20} \, \rm cm^{-2})$ &  $\Gamma$ & $F_{\rm 1keV} \, (\mu \rm Jy)$    &  $\chi^2_{\rm red}$ (d.o.f.)\\
\hline
July 29--31  &  $5.02 \pm 0.11$ &  $2.565 \pm 0.006$  &  $3.260 \pm 0.016$ &  1.09 (2671)\\
August 2--4  &  $5.20 \pm 0.11$ &  $2.564 \pm 0.006$  &  $2.995 \pm 0.015$ &  1.08 (2787)\\
August 4--6  &  $5.04 \pm 0.10$ &  $2.560 \pm 0.006$  &  $3.010 \pm 0.015$ &  1.05 (2832)\\
August 8--10 &  $4.57 \pm 0.11$ &  $2.551 \pm 0.006$  &  $2.693 \pm 0.015$ &  1.00 (2657)\\
August 16--17&  $4.92 \pm 0.09$ &  $2.594 \pm 0.005$  &  $3.678 \pm 0.018$ &  1.08 (2865)\\
Aug~30-Sep~1&  $5.65 \pm 0.10$ &  $2.595 \pm 0.005$  &  $3.912 \pm 0.017$ &  1.08 (2948)\\
\hline
\end{tabular}
\end{table*}

The power-law fits are statistically better, but suggest an absorption that is 23--52\% higher than the Galactic value. This may indicate a spectral curvature that, according to the log-parabolic fits, is however not very pronounced. Indeed, the $\beta$ parameter ranges between 0.06 on August~8--10 and 0.15 on August~30--September~1. We note that the minimum/maximum curvature corresponds to the minimum/maximum flux density, respectively.

\begin{table*}
\caption{Results of the EPIC spectral fitting with a log-parabolic model with absorption fixed at the Galactic value $N_{\rm H} =3.72 \times 10^{20} \rm \, cm^{-2}$.}
\label{epic_logpar}
\begin{tabular}{lllll}
\hline
Date &  $\alpha$ &  $\beta$ & $F_{\rm 1keV} \, (\mu \rm Jy)$    &  $\chi^2_{\rm red}$ (d.o.f.)\\
\hline
July 29--31  &  $2.487 \pm 0.004$ &  $0.104 \pm 0.010$  &  $3.179 \pm 0.012$ &  1.13 (2671)\\
August 2--4  &  $2.476 \pm 0.004$ &  $0.115 \pm 0.010$  &  $2.909 \pm 0.011$ &  1.14 (2787)\\
August 4--6  &  $2.482 \pm 0.003$ &  $0.101 \pm 0.009$  &  $2.931 \pm 0.012$ &  1.10 (2832)\\
August 8--10 &  $2.501 \pm 0.004$ &  $0.057 \pm 0.010$  &  $2.645 \pm 0.013$ &  1.04 (2657)\\
August 16--17&  $2.521 \pm 0.003$ &  $0.102 \pm 0.008$  &  $3.596 \pm 0.016$ &  1.12 (2865)\\
Aug~30--Sep~1&  $2.481 \pm 0.003$ &  $0.151 \pm 0.009$  &  $3.767 \pm 0.013$ &  1.18 (2948)\\
\hline
\end{tabular}
\end{table*}

\begin{figure}
   \centering
   \resizebox{\hsize}{!}{\includegraphics[angle=-90]{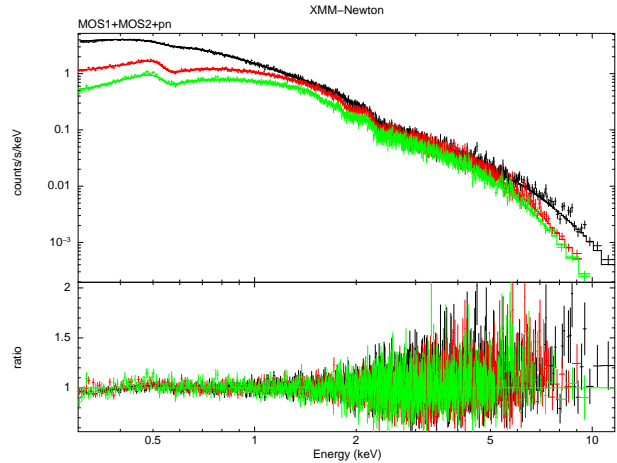}}
   \caption{X-ray spectra of PG 1553+113 acquired by the EPIC instruments onboard {\em XMM-Newton} on 2015 August 8--10. Black, red and green symbols represent MOS1, MOS2 and pn data, respectively. The folded model, an absorbed log parabola with $N_{\rm H} =3.72 \times 10^{20} \rm \, cm^{-2}$, is shown by solid lines of the same colour. The ratio between the data and the folded model is plotted in the bottom panel.}
    \label{xmmspec}
    \end{figure}

For illustrative purposes, Fig.\ \ref{xmmspec} shows the three EPIC spectra acquired on August 8--10, with the absorbed log-parabolic model folded and data-to-model ratio.

\subsubsection{Light curves}
\begin{figure}
    \vspace{0.5cm}
    \centerline{
    \psfig{figure=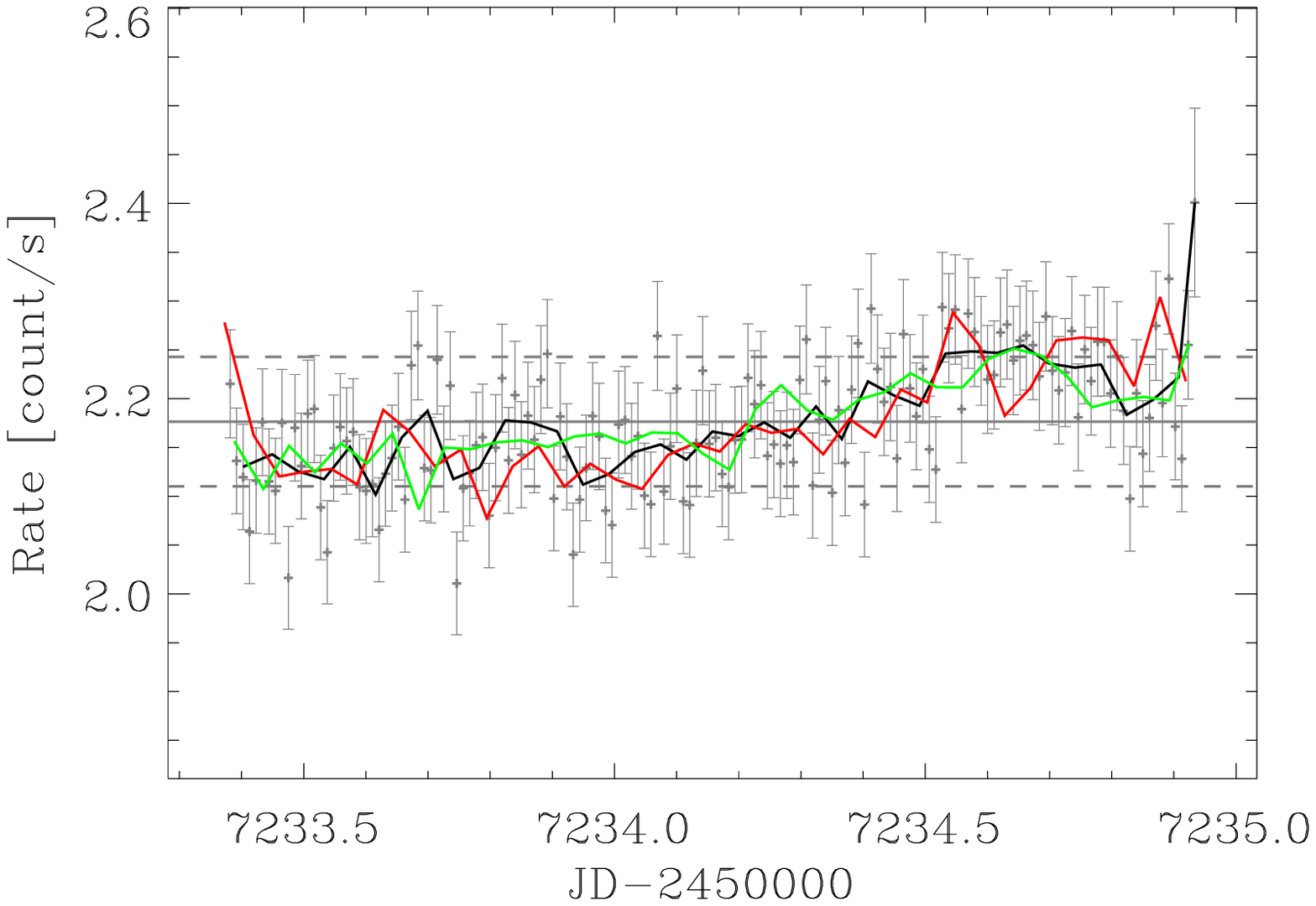,width=0.50\linewidth}
    \psfig{figure=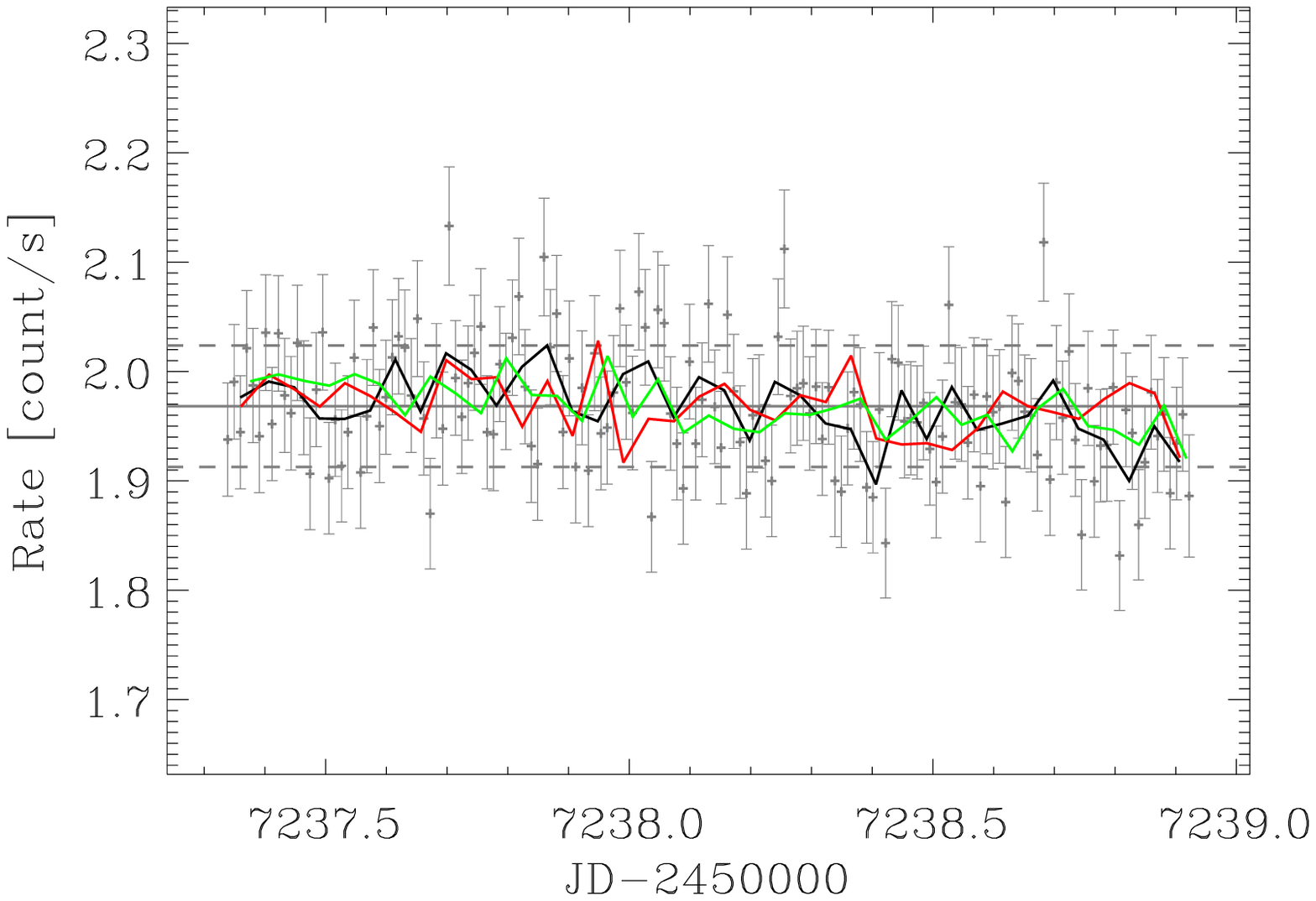,width=0.50\linewidth}
    }
    \vspace{0.5cm}
    \centerline{
    \psfig{figure=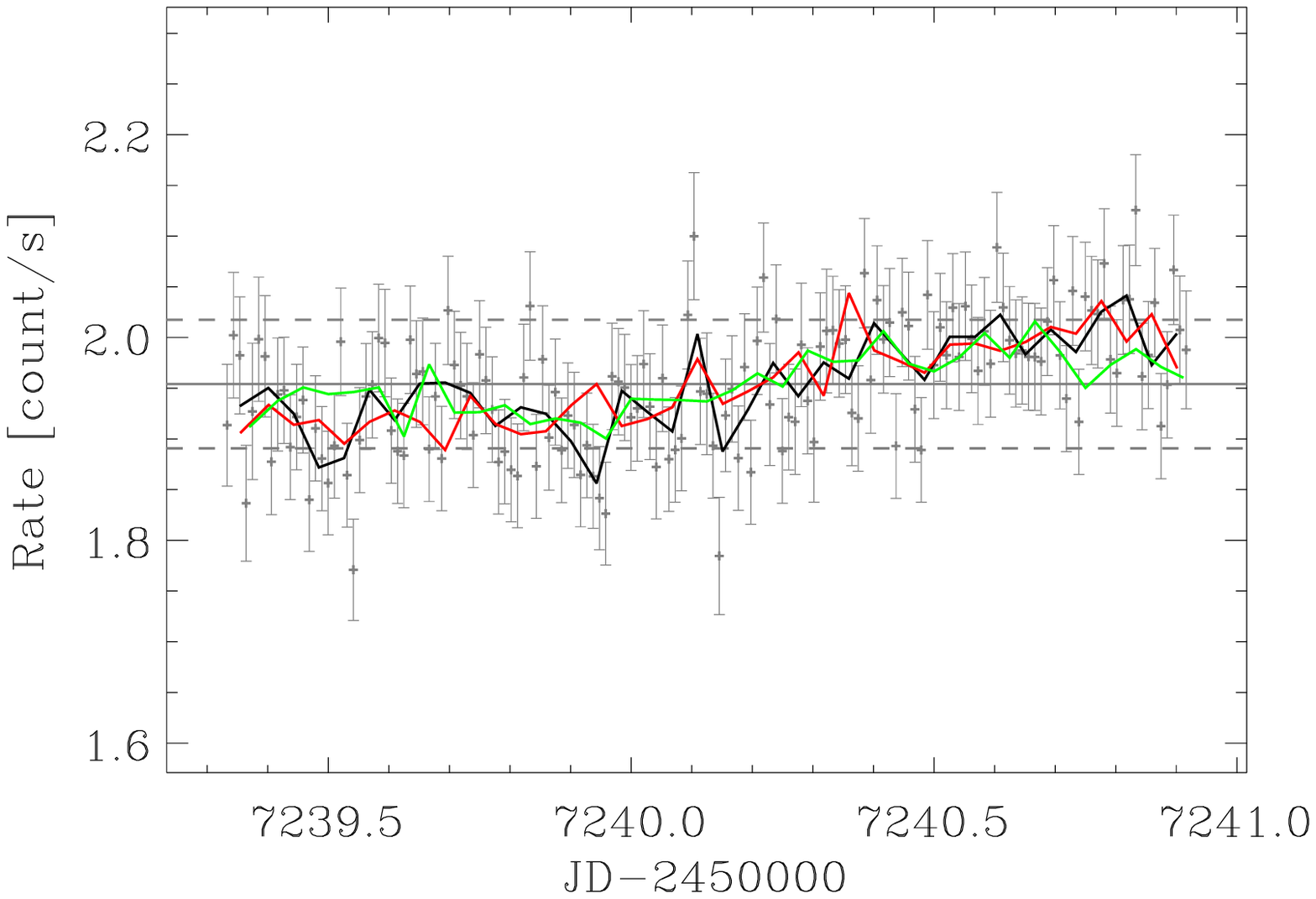,width=0.50\linewidth}
    \psfig{figure=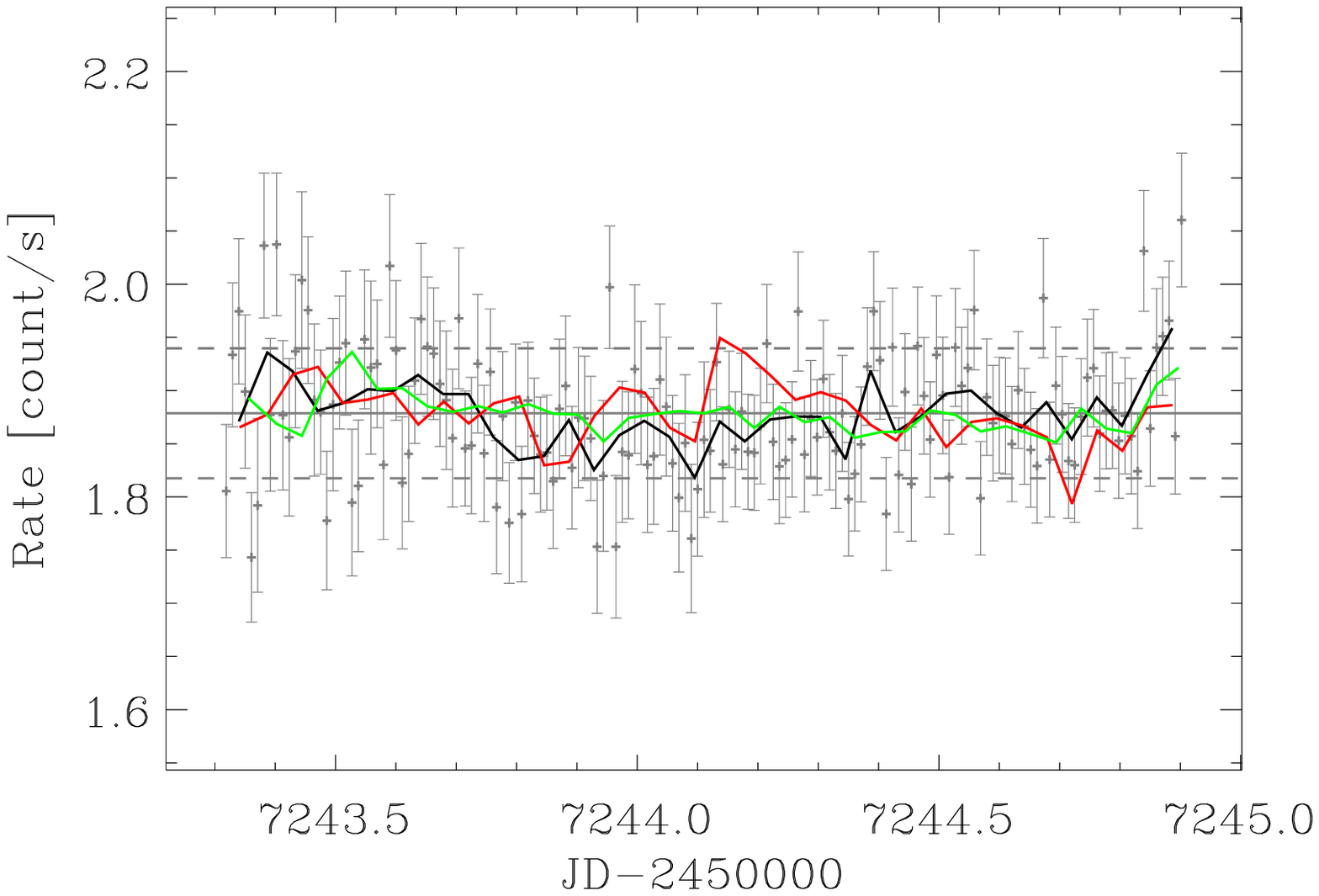,width=0.50\linewidth}
    }
    \vspace{0.5cm}
    \centerline{
    \psfig{figure=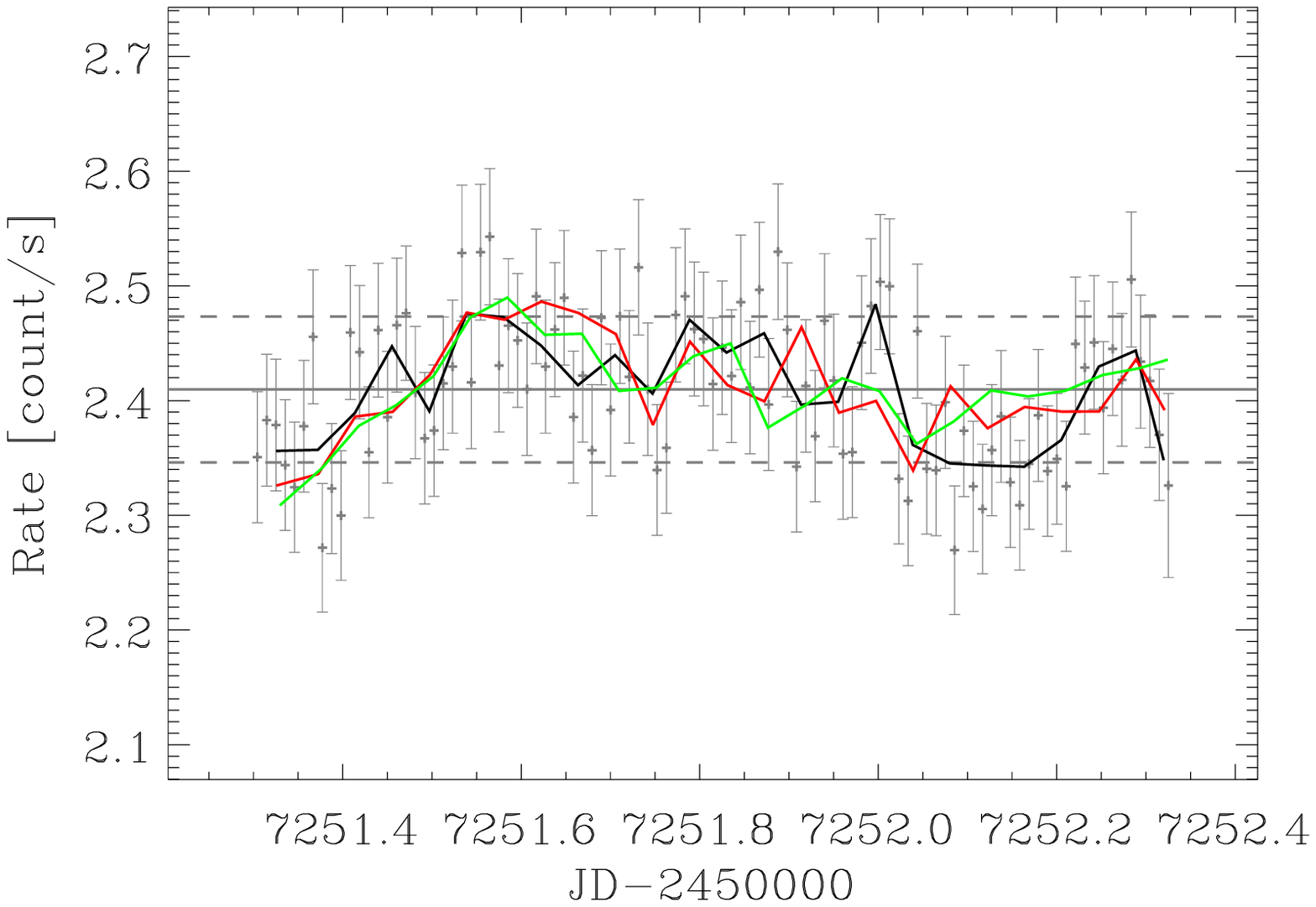,width=0.50\linewidth}
    \psfig{figure=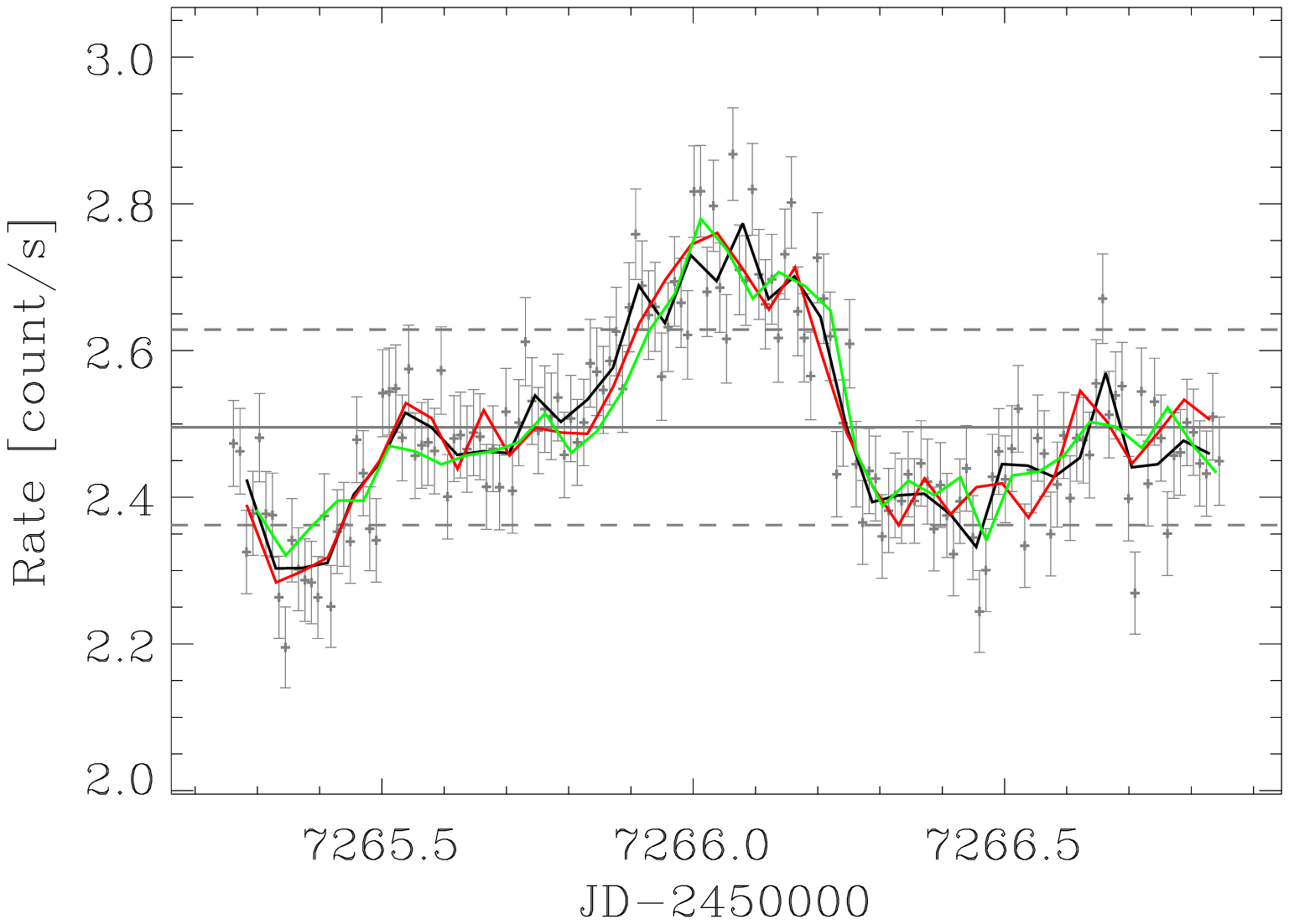,width=0.50\linewidth}
    }
     \caption{X-ray background-corrected light curves from the EPIC detectors during the {\em XMM-Newton} pointings. Grey plus signs correspond to the MOS1 rates obtained with a binning of 900 s, while the black line shows the result of a binning over 3600 s. The red and green lines represent 3600 s binning on the MOS2 and pn data, respectively, properly scaled so that the mean values of the three datasets match.}
    \label{varix}
   \end{figure}

We checked the EPIC data for possible fast variability, which can put constraints on the size of the emitting region. 
We built MOS1, MOS2, and pn light curves following standard prescriptions\footnote{http://www.cosmos.esa.int/web/xmm-newton/sas-thread-timing}. We performed a barycentric correction on the event files with the {\tt barycen} task. Source and background light curves were extracted in the 0.2--10 keV energy range from the same regions used for the spectral analysis. For the pn, we considered single and double events only, while for the MOS we also included triples and quadruples. The task {\tt epiclccorr} was finally run to obtain background-subtracted light curves, after performing both absolute (energy-dependent) and relative (time-dependent) corrections.
Figure \ref{varix} shows the corrected light curves.
Some variability can be recognized especially during the last {\em XMM-Newton} pointing, on a hourly time scale.
This suggests that the size of the emitting region is $R \la \delta \times 10^{14} \, \rm cm$, where $\delta$ is the Doppler beaming factor and typically ranges between 4 and 35 in blazars \citep{sav10}.

\subsection{OM}
During each XMM-Newton observation, the Optical Monitor \citep[OM;][]{mas01} instrument acquired many exposures in the optical and UV wide-band filters $V$, $B$, $U$, $W1$, $M2$ and $W2$. For each filter one exposure in imaging mode (5000 s long) is followed by one exposure in fast, imaging mode (usually 4400 s long); the sequence ends with a number (4--14) of fast exposures in $W2$ band, for a total integration time of 6--17 hours in this band.

We processed the data with the pipelines {\tt omichain} and {\tt omfchain} of the {\sevensize \bf SAS} package, and analysed the results with the {\tt omsource} task.
Source magnitudes corresponding to the imaging mode are shown in Fig.\ \ref{uvot}.
Flux densities were obtained by multiplying the count rates for the conversion factors derived from observations of white dwarf standard stars\footnote{http://xmm.esac.esa.int/sas/8.0.0/watchout/}.
We conservatively assumed a 10\% uncertainty on the count rates for all filters.

\subsection{Joint EPIC-OM spectral fitting}

The exceptional quality of the {\em XMM-Newton} data encouraged us to try a joint fit of the EPIC and OM data. 
We thus used the {\sevensize \bf  SAS} task {\tt om2pha} to convert the OM data into ``spectra" that can be read into {\sevensize \bf  Xspec}.
The fitting model must be curved, since we are considering a broad frequency range that goes through the synchrotron peak.
We adopted a log-parabolic model \citep{lan86,mas04}, with reddening applied to the OM data and absorption to the EPIC data (redden*tbabs*logpar). Galactic reddening was set to $E(B-V)=0.05$ \citep[see][]{rai15}. 
Fixing both $E(B-V)$ and $N_{\rm H}$ to their Galactic values results in a poor fit, which in general improves significantly when thawing the reddening parameter, but not so much when thawing $N_{\rm H}$. 
Leaving both parameters free to vary gives an even slightly better fit, but its goodness is still questionable ($\chi^2/\nu=1.3$--2.4, see Table \ref{joint}). 
In general, a free $E(B-V)$ gives a value smaller than the Galactic one, while a free $N_{\rm H}$ goes in the opposite direction, both results suggesting a gas/dust ratio $r=N_{\rm H}/E(B-V)$ higher than the average value in the Milky Way, $r_{\rm MW}$. Assuming $r_{\rm MW}=4.93 \times 10^{21} \, \rm cm^{-2} \, mag^{-1}$ from \citet{dip94}, we obtain gas/dust ratios from 1.5 to 2.6 times the Galactic value.
The other parameters of the model, i.e.\ $\alpha, \beta$ and the normalization, do not change much when freezing or thawing reddening and absorption.

We conclude that the log-parabolic model does not give a statistically acceptable description of the OM+EPIC spectrum. The above attempts 
indicate that the OM values for the source brightness are too high to allow for a fair connection between the optical--UV and X-ray spectra.
However, we have to consider that these joint fits are strongly constrained by the large number of degrees of freedom (d.o.f.) of the X-ray spectra that give a heavy weight to the X-ray data.
A comparison with optical--UV and X-ray data by {\em Swift} as well as with the WEBT data from the ground is crucial to assess whether the low goodness of the joint fits is due to data uncertainties or if it has to do with the source spectral properties.

\begin{table*}
\caption{Results of the joint OM+EPIC spectral fitting with a log-parabolic model with free reddening and absorption.}
\label{joint}
\begin{tabular}{lllll}
\hline
Date & $E(B-V) \, (10^{-2} \, \rm mag)$ &  $N_{\rm H} \,  (10^{20} \, \rm cm^{-2})$ &  $\chi^2_{\rm red}$ (d.o.f.) & $r/r_{\rm MW}$\\
\hline
July 29--31  &  $3.32 \pm 0.07$ &  $4.30 \pm 0.05$  &  2.36 (2669) & 2.6\\
August 2--4  &  $3.74 \pm 0.08$ &  $4.40 \pm 0.05$  &  1.51 (2786) & 2.4\\
August 4--6  &  $3.49 \pm 0.08$ &  $4.27 \pm 0.05$  &  1.75 (2830) & 2.5\\
August 8--10 &  $2.88 \pm 0.09$ &  $3.56 \pm 0.05$  &  1.71 (2655) & 2.5\\ 
August 16--17&  $5.02 \pm 0.09$ &  $3.81 \pm 0.04$  &  1.32 (2859) & 1.5\\  
Aug~30--Sep~1&  $4.20 \pm 0.08$ &  $4.47 \pm 0.05$  &  1.52 (2947) & 2.2\\   
\hline
\end{tabular}
\end{table*}

\section{Observations by {\em Swift}}
The {\em Swift} satellite carries three instruments: the Ultraviolet/Optical telescope \citep[UVOT;][]{rom05}, the X-ray Telescope \citep[XRT;][]{bur05}, and the BAT telescope.
Data were reduced with the {\sevensize \bf  HEAsoft}\footnote{http://heasarc.nasa.gov/lheasoft/} package version 6.18 available at the NASA's High Energy Astrophysics Science Archive Research Center (HEASARC).

\subsection{XRT}
We reduced the XRT data with the CALDB calibration release 20160121. 
The task {\tt xrtpipeline} was run on the observations performed in pointing mode with standard screening criteria, selecting events with grades 0--12 for the PC mode and grades 0--2 for the WT mode. We retained all the PC and WT observations with exposure time greater than 1 min and 30 sec, respectively.
Data acquired in PC mode were affected by pile-up, so we extracted the source counts from an annular region with inner and outer radius of 10 and 75 arcsec, respectively, while the background counts were derived from an annulus with radii of 100 and 150 arcsec.
For the observations performed in WT mode, we extracted the source counts in a circular region with 70 arcsec radius, and the background counts in the same annulus used for the PC mode.
The task {\tt xrtmkarf} was used to create ancillary response files to correct for
hot columns, bad pixels, and the loss of counts caused by using an annular extraction region in the pile-up case.
We binned each spectrum and associated it with the corresponding background, redistribution matrix (rmf), and ancillary (arf) files with the task {\tt grppha}, setting a binning of at least 20 counts for each spectral channel in order to use the $\chi^2$ statistic.
We fitted the XRT spectra with the same models adopted for the EPIC spectra (see Sect.\ \ref{epic}).
When the number of degrees of freedom was less or equal to 10 we checked the results with the Cash's $C$ statistic.
The X-ray light curves will be presented and discussed in Sect.\ \ref{long}.

   \begin{figure}
   \centering
   \resizebox{\hsize}{!}{\includegraphics{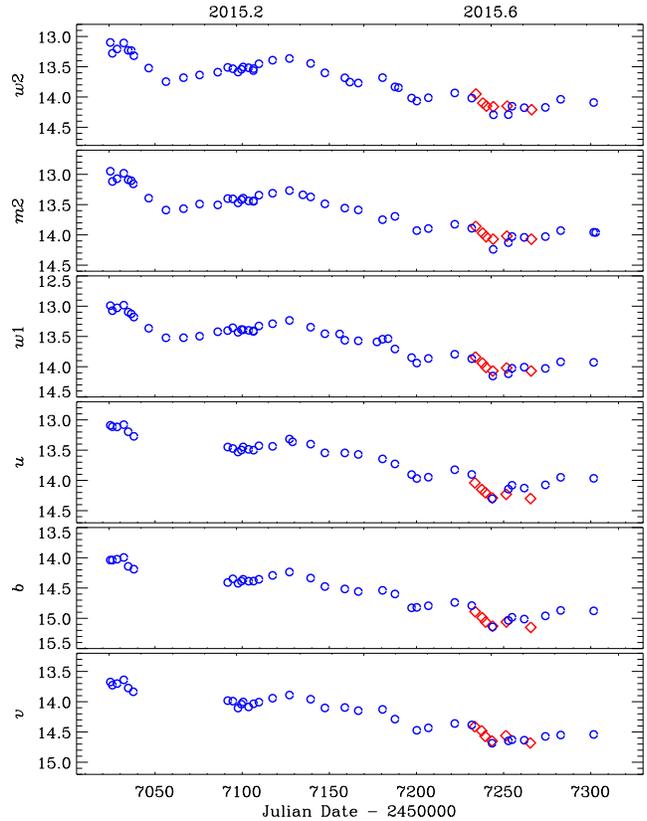}}
   \caption{UV--optical observed magnitudes obtained from {\em Swift}-UVOT data (blue circles) and {\em XMM-Newton}-OM observations in imaging mode (red diamonds).
Errors are smaller than the symbol size.}
    \label{uvot}
    \end{figure}

\subsection{UVOT}
UVOT observed PG~1553+113 with the $v$, $b$ and $u$ optical filters, and $w1$, $m2$ and $w2$ UV filters.
Data were processed with the calibration release 20160321 of the CALDB data base accessible from the HEASARC. For each epoch, 
we summed multiple images in the same filter with the task {\tt uvotimsum} and then performed aperture photometry with {\tt uvotsource}. 
Source counts were extracted from a circular region with 5 arcsec radius centred on the source, while background counts were obtained from a circle with 20 arcsec radius located in a source-free field region close to the source.
The results were checked against the results of aperture photometry on the single images. We also investigated the occurrence of small scale sensitivity effects\footnote{http://swift.gsfc.nasa.gov/analysis/uvot\textunderscore digest/sss\textunderscore check.html} due to the possibility that the source falls on the small areas of the detector with lower throughput.

Figure \ref{uvot} displays the optical and UV magnitudes derived from the UVOT data as well as those coming from
the OM observations in imaging mode. 
The comparison between the results of the two instruments shows a fair agreement, as the OM data points lie on the path traced by the UVOT data points, and the differences between simultaneous data are within the uncertainties.
The light curves present a general decreasing trend, with the {\em XMM-Newton} observations occurring when the source was in its faintest state.

De-reddened flux densities were obtained from UVOT data by following the prescriptions of \citet{rai15}.  
\begin{table}
\caption{The optical, near-IR and radio observatories participating in the 2015 WEBT campaign on PG~1553+113.}
\label{obs}
\begin{tabular}{lll}
\hline
Observatory & Country & Bands \\
\hline
\multicolumn{3}{c}{\it Optical} \\
Abastumani & Georgia & $R$ \\
AstroCamp & Spain & $R$ \\
Belogradchik & Bulgaria & $BVRI$ \\
Catania & Italy & $BVRI$ \\
Crimean$^1$ & Russia & $BVRI$ \\
Lulin$^1$ & Taiwan & $g,r,i$ \\
Michael Adrian & Germany & $RI$ \\
Mt.~Maidanak$^2$ & Uzbekistan & $BVRI$ \\
New Mexico Skies & USA & $R$ \\
ROVOR & USA & $RV$ \\
Rozhen$^2$ & Bulgaria & $BVRI$ \\
Siding Spring & Australia & $R$ \\
Sirio & Italy & $R$ \\
Skinakas$^1$ & Greece & $BVRI$ \\
St.~Petersburg$^1$ & Russia & $BVRI$ \\
Teide$^2$ & Spain & $VR$ \\
Tijarafe & Spain & $BR$ \\
ASV$^3$ & Serbia & $BVRI$ \\
\hline
\multicolumn{3}{c}{\it Near-infrared} \\
Campo Imperatore & Italy & $JHK$ \\
Roque de los Muchachos (TNG) & Spain & $JHK$ \\
Teide & Spain & $JHK$ \\
\hline
\multicolumn{3}{c}{\it Radio} \\
Medicina & Italy & 5, 8 GHz \\
Mets\"ahovi & Finland & 37 GHz \\
\hline
$^1$ photometry and photopolarimetry\\
$^2$ two telescopes\\
$^3$ Astronomical Station Vidojevica
\end{tabular}
\end{table}

   \begin{figure*}
   \centering
   \includegraphics[width=10cm]{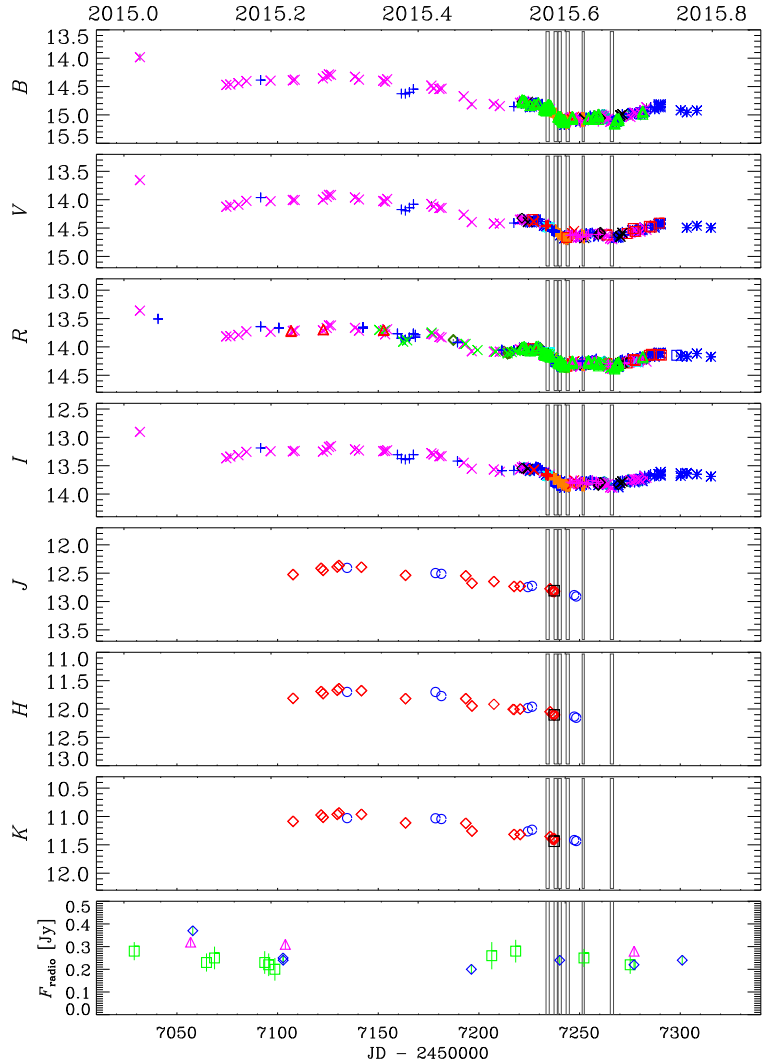}
   \caption{Final $BVRIJHK$ light curves of PG 1553+113 obtained by the WEBT collaboration. Different colours and symbols highlight data from different telescopes. Grey boxes indicate the {\em XMM-Newton} observations. In the bottom panel flux densities at 5 GHz (purple triangles), 8 GHz (blue diamonds), and 37 GHz (green squares) are shown.}
    \label{webt}
    \end{figure*}

   \begin{figure*}
   \centering
   \includegraphics[width=12cm]{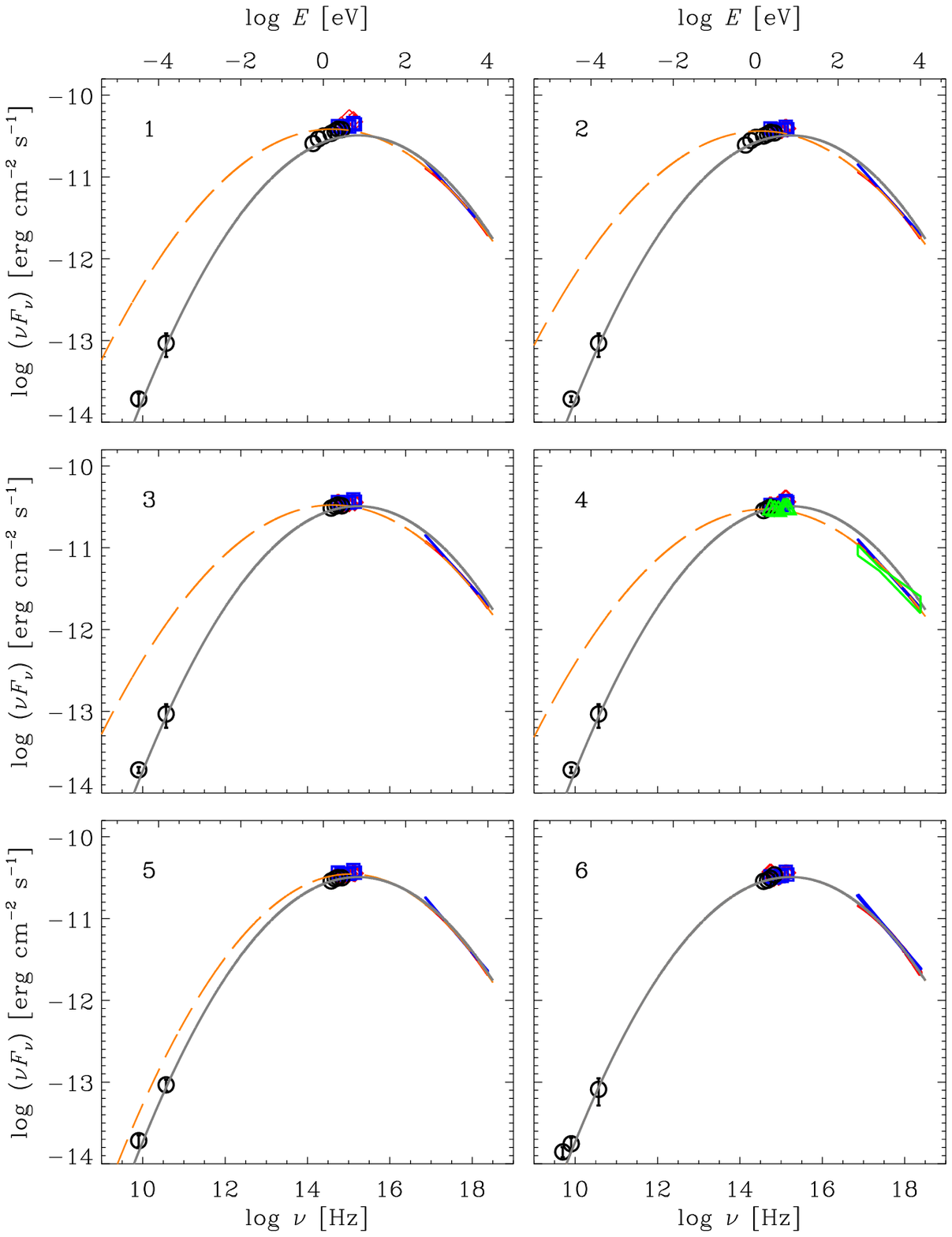}
   \caption{SEDs of PG~1553+113 during the six {\em XMM-Newton} pointings in July--September 2015. Both the power-law (blue) and log-parabolic (red) fits to the MOS1, MOS2, and pn spectra analysed together are shown. The OM data acquired in imaging mode are plotted as blue squares, while red diamonds display data taken in fast, imaging mode. Black circles represent WEBT data points. On August 8--10 a simultaneous {\em Swift} observation was performed; its results are plotted in green. The grey continuous line represents the log-parabolic fit to the joint OM+EPIC data of the last {\em XMM-Newton} pointing; it is reported in all panels to guide the eye. The log-parabolic fits to the joint OM+EPIC data of each epoch are shown as orange long-dashed lines.}
    \label{sed}
    \end{figure*}

\section{WEBT data}

The WEBT campaign on PG~1553+113 in 2015 was motivated by and focussed on the giant amount of {\em XMM-Newton} time awarded in July--September.
Nonetheless, we collected data from the whole optical observing season.
Table \ref{obs} lists the participating observatories, together with their country and observing bands.

In the optical band, 21 telescopes in 18 observatories acquired data for the campaign.
The calibration of the source magnitude was obtained with respect to the photometric sequence published by \citet{rai15}.
We collected 3120 data points in the Johnson-Cousins' $B, V, R, I$ bands\footnote{Lulin's data were acquired in the SDSS $g, r, i$ filters and transformed into $B, V, R, I$ values with the \citet{cho08} calibrations.}, which were carefully assembled to obtain homogeneous and reliable light curves. In particular, overlapping of different datasets sometimes revealed offsets that were corrected by shifting deviant datasets to make them match the common trend traced by the others. Moreover, the scatter of dense but noisy datasets was reduced either by binning data points close in time or by deleting points that lie more than 1 standard deviation out of the night average, if no trend is recognizable. Finally, we discarded data with large errors (larger than 0.1 mag) or standing clearly out of the location defined by the bulk of the other points. The cleaning process is facilitated by the simultaneous inspection of the multifrequency light curves.
The final light curves are displayed in Fig.\ \ref{webt}. They show the same decreasing trend already seen in the {\em Swift}-UVOT light curves in Fig.\ \ref{uvot}. The sampling is denser during the core of the campaign, centred on the {\em XMM-Newton} observations.

Near-infrared observations were performed at the Campo Imperatore, Roque de los Muchachos (TNG), and Teide Observatories. 
Data reduction and analysis procedures are described in \citet{rai14}.
The $JHK$ flux densities are plotted in Fig.\ \ref{webt} and show a course similar to that observed in the optical bands.

Radio data were acquired at 5 GHz and 8 GHz with the 32-m dish of the Medicina Radio Astronomical Station and at 37 GHz with the 14-m radio telescope of the Mets\"ahovi Radio Observatory. Data are processed as explained in \citet{dam12} and \citet{ter98}. The radio flux variability is modest.

   \begin{figure}
   \centering
   \resizebox{\hsize}{!}{\includegraphics{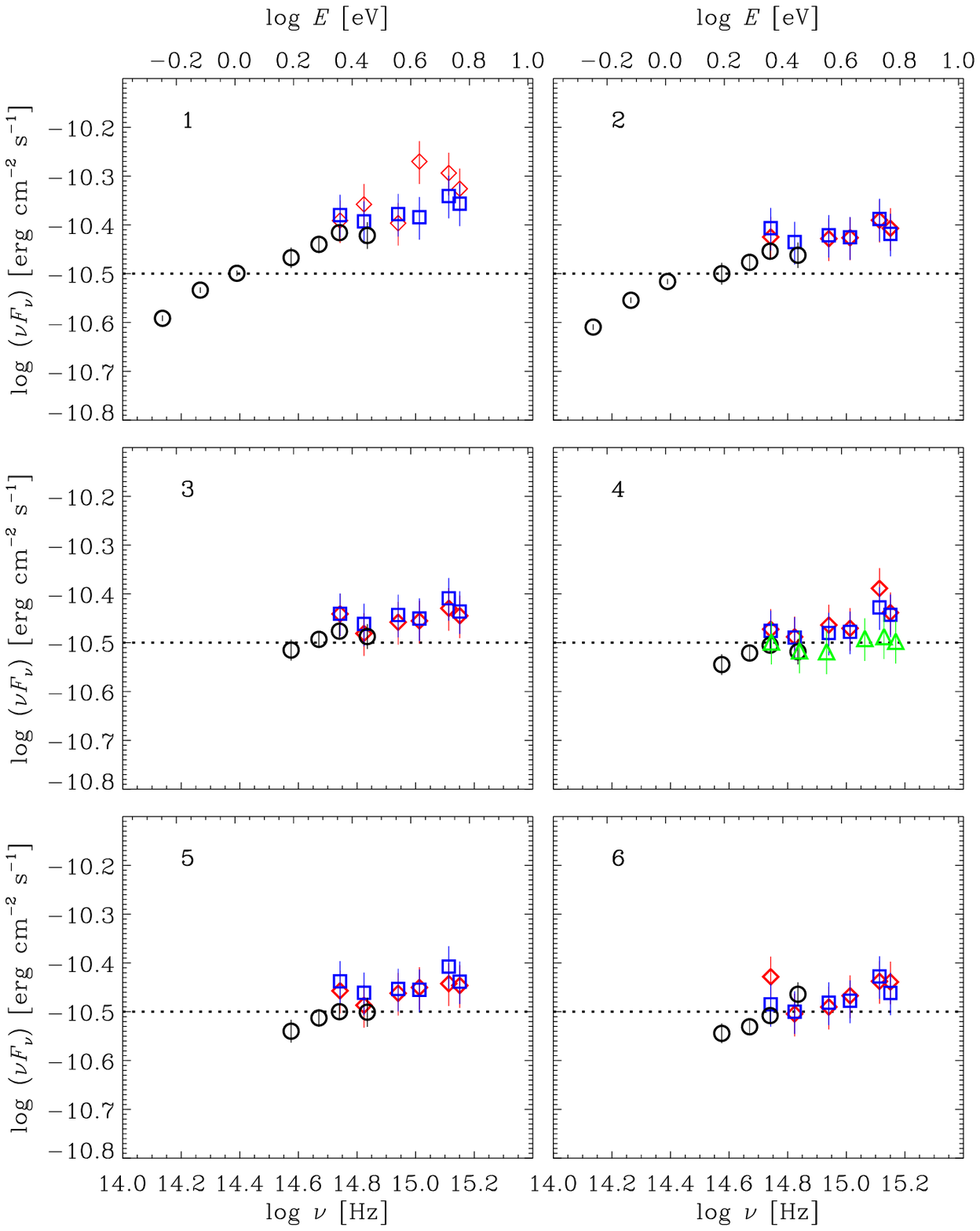}}
   \caption{A zoom into the near-IR-to-UV frequency range of the SEDs reported in Fig.\ \ref{sed}. The dotted line is drawn to guide the eye through the flux changes.}
    \label{sed_zoom}
    \end{figure}

\section{Spectral energy distribution}

Figure \ref{sed} shows the broad-band SEDs of the source during the six {\em XMM-Newton} pointings.
In X-rays, both fits to the EPIC data (power law with free absorption and log-parabola with Galactic absorption) are plotted. 
A zoom into the near-infrared-to-UV frequency range is displayed in Fig.\ \ref{sed_zoom}.
In both figures imaging and fast, imaging OM data are plotted separately.

The August 8--10 SED also includes the {\em Swift} data obtained simultaneously to the {\em XMM-Newton} pointing. 
They indicate a slightly lower flux in the UV and a bit harder X-ray spectrum than the {\em XMM-Newton} data.
The UVOT spectrum also shows a less pronounced curvature with respect to those analysed by \citet{rai15}.

Simultaneous optical and near-infrared data taken during the WEBT campaign are added.
Flux densities were corrected for Galactic absorption according to the prescriptions given by \citet{rai15}. 
The WEBT data are in agreement, within errors, with the space data in the overlapping spectral range.
In the radio band we plotted the nearest data in time, with a tolerance of a couple of weeks.

In each panel we show the result of the log-parabolic fit to the joint OM and EPIC data of each epoch, together with that of the sixth {\em XMM-Newton} pointing (see Section 2.3).
While in the first five epochs the joint OM+EPIC fits overproduce the WEBT data (radio+near-infrared in the first two cases, radio only in the other three), the fit obtained in the sixth epoch also matches the contemporaneous radio data and suggests that the synchrotron peak lies in the near-UV band.
It is also consistent with the broad-band SED of the fifth pointing, while slightly overproduces the X-ray flux of the other epochs and slightly underproduces the UV flux of the first one. 

\begin{figure}
   \centering
   \resizebox{\hsize}{!}{\includegraphics{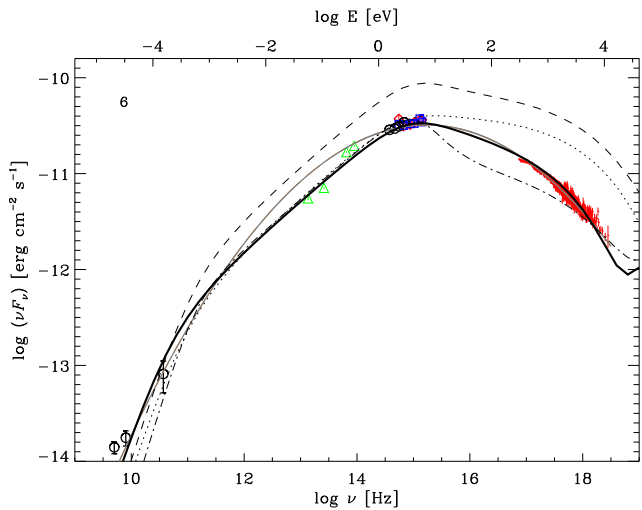}}
   \caption{SEDs of PG~1553+113 during the sixth {\em XMM-Newton} pointing of 2015 August 30 -- September 1. Symbols are as in Fig.\ \ref{sed}. The EPIC-pn spectrum obtained with a log-parabolic fit is shown in red. Infrared data from the AllWISE Source Catalog are plotted as green triangles. The grey continuous line represents the log-parabolic fit to the joint OM+EPIC data. The dashed, dotted, and dot-dashed lines show the three fits to high, intermediate, and low X-ray states of the source obtained by \citet{rai15} with an inhomogeneous SSC helical jet model. The black solid line represents a fit to the 2015 August 30 -- September 1 SED with the same model.}
    \label{sed_fab6}
    \end{figure}

In Fig.\ \ref{sed_fab6} we compare this log-parabolic model with the fits to a high, intermediate, and low X-ray state of the source obtained by \citet{rai15} with an inhomogeneous SSC helical jet model \citep{vil99,rai09}. The 2015 UV-to-X-ray flux is in between those characterizing the intermediate and low states, and the X-ray spectrum appears softer, while the radio flux is higher.  
We also show a realization of the same jet model that can satisfactorily reproduce the SED of 2015 August 30 --September 1. 
With respect to the \citet{rai15} fits, it was obtained by changing the three parameters defining the jet orientation, and by lowering the emitted minimum and maximum frequencies at the jet apex. The new parameters are $a=75\degr$, $\psi=24\degr$, $\phi=18\degr$, and $\log \nu'=15.4$--18.2 (rest frame). All the other model parameters are fixed to the values used in \citet{rai15}.
The largest separation between the log-parabolic and the jet model fits occurs in the infrared, so in Fig.\ \ref{sed_fab6} we added the WISE data downloaded from the AllWISE Source Catalog\footnote{http://irsa.ipac.caltech.edu/cgi-bin/Gator/nph-dd}. Although the WISE data are not contemporaneous with the other data, they suggest a power-law shape of the SED in that wavelength range, more in agreement with the jet model.

In conclusion, the 2015 SEDs indicate an intermediate UV-to-X-ray state as compared to those analysed by \citet{rai15}. Moreover, the smaller curvature of the UV and X-ray spectra in 2015 allows for a better connection between the UV and X-ray spectra, so that in first approximation a simple log-parabolic fit appears suitable to describe at least the SEDs with the highest X-ray fluxes.
However, a more detailed analysis requires a more complex model, like the inhomogeneous SSC helical jet model that has already been applied to previous, more problematic SEDs of this source.

\section{Long-term photometric and polarimetric behaviour}
\label{long}

\begin{figure*}
   \centering
   \includegraphics[width=12cm]{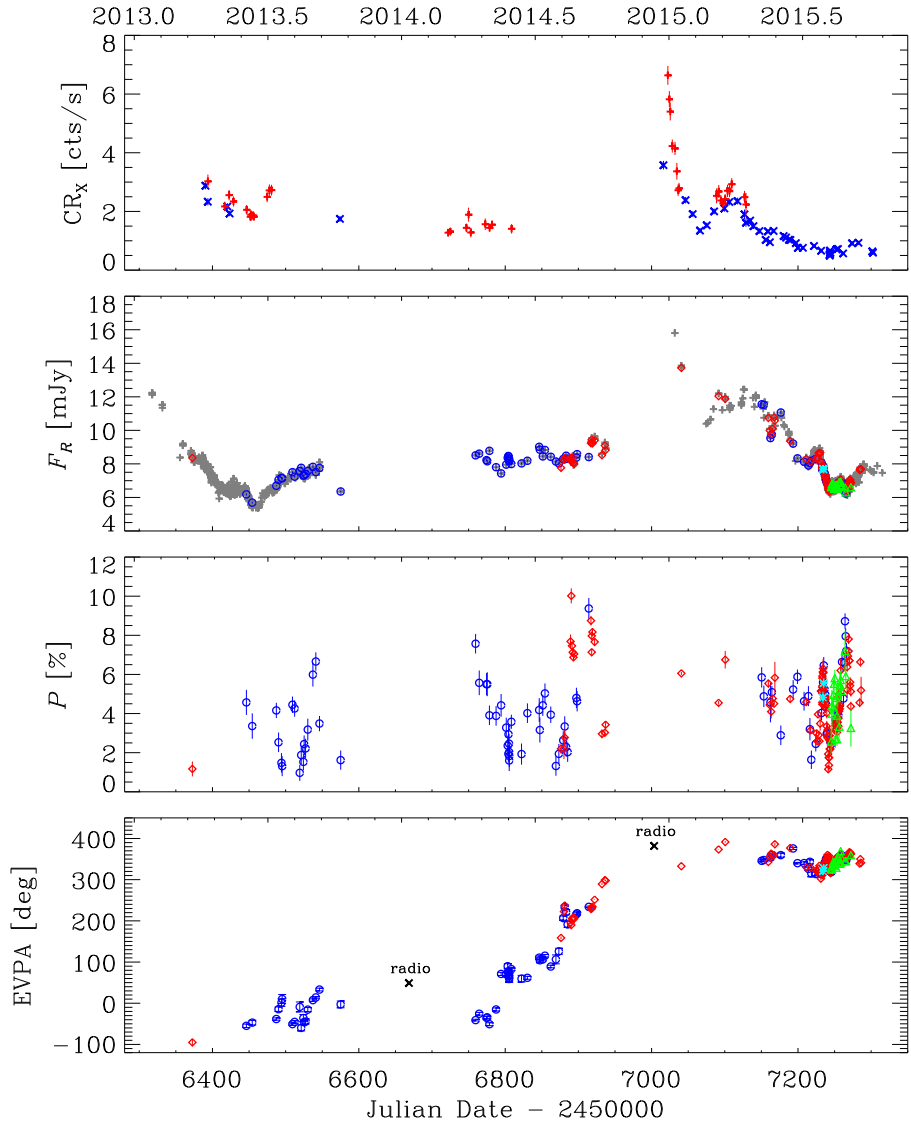}
   \caption{From top to bottom: time trend of the X-ray count rate (top panel), de-absorbed flux density in $R$ band, percentage polarization $P$, and electric vector polarization angle EVPA. In the top panel, red plus signs indicate {\em Swift}-XRT observations performed in WT mode, while blue crosses those in PC mode.
In the other panels, red diamonds, cyan asterisks, blue circles and green triangles refer to data from the Crimean, Lulin, Skinakas, and St.~Petersburg observatories, respectively. Grey plus signs in the second panel indicate the whole WEBT dataset in the $R$ band, while the black crosses in the last panel 
correspond to VLBA measurements at 15 GHz obtained by the MOJAVE Project. }
    \label{pola}
    \end{figure*}

\begin{figure*}
   \centering
   \includegraphics[width=10cm]{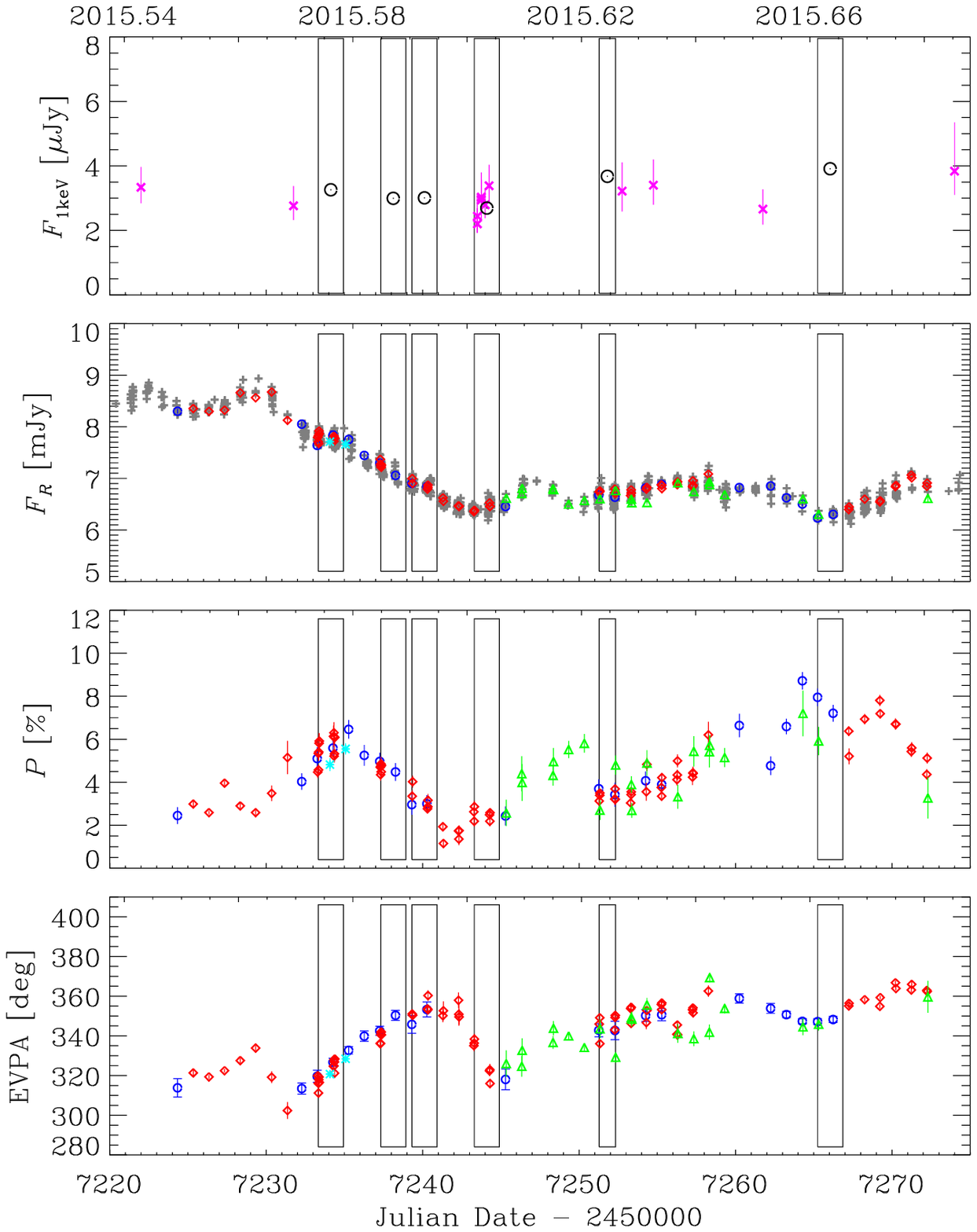}
   \caption{A zoom into the trends shown in Fig.\ \ref{pola}. Boxes indicate the epochs of the {\em XMM-Newton} pointings.
In the top panel, the 1 keV flux density is given for both the {\em Swift}-XRT observations (purple crosses) and the {\em XMM-Newton}-EPIC ones (black empty circles).
They have been obtained by applying an absorbed power-law model with free $N_{\rm H}$.}
    \label{zoom}
    \end{figure*}

The long-term (2013--2015) optical and X-ray light curves are presented in Fig.\ \ref{pola} and compared with the behaviour of the polarization percentage $P$ and of the electric vector position angle (EVPA). Figure \ref{zoom} shows an enlargement of the period including the {\em XMM-Newton} pointings examined in this paper and the corresponding WEBT campaign. 

A prominent flare was detected in the X-ray band at the very end of 2014--very beginning of 2015, which was characterised by a rapid fall and was followed by a minor flare. 
The ratio between the maximum and the minimum flux is $\sim 13$.
The decreasing part of the X-ray major flare was observed also in the optical band, where it was probably more contained. The maximum observed flux density in the $R$ band was only $\sim 2.6$ times higher than the minimum in 2015. The optical counterpart of the X-ray second, minor flare appears to be more extended in time.

\begin{figure}
\includegraphics[width=84mm]{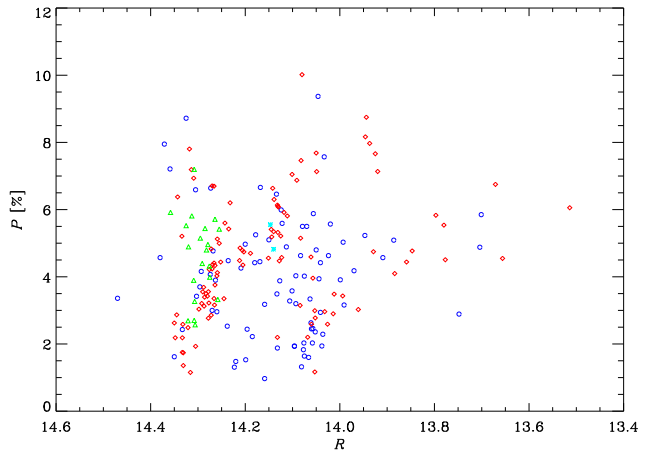}
\caption{Polarization percentage in the $R$ band as a function of the $R$-band magnitude.}
\label{permag}
\end{figure}

Optical polarimetric data were acquired at the Crimean, Lulin, Skinakas, and St.~Petersburg observatories. Two points with large error ($\sigma_P/P>2$) were removed.
The $\pm n \, \pi$ ($n \in N$) ambiguity on the EVPA was treated by minimizing the difference between adjacent points.
The value of $P$ rapidly oscillates between $\sim 1 \%$ and $\sim 10 \%$ in a way that does not appear to be correlated with the optical (see also Fig.\ \ref{permag}) or X-ray brightness behaviour. The first polarimetric datum in 2015, when the source flux was still high, does not reveal particularly high polarization. Instead, $P$ reached the highest values in 2014, when the flux was rather stable and relatively low.

The EVPA is quite variable too. It increases during all the 2014 observing season and makes a rotation of 350\degr\ in about six months, with a mean rate of $\sim 60\degr$/month, while it remains more or less stable after the major flare. Notice however that the jump at JD=2456794 is $\sim 86 \degr$, i.e.\ very close to $90 \degr$. Therefore at that epoch the EVPA minimizing procedure sets the rotation counter-clockwise (increasing EVPA) instead of clockwise (decreasing EVPA) because of a difference of a few degrees only. This is of the same order of the overall error on the two data points defining the EVPA jump. The opposite choice would produce a smaller total rotation and rate, about 230\degr\ in 4.8 months, i.e.\ $\sim 50\degr$/month\footnote{For a discussion on the treatment of the EVPA $\pm n \, \pi$ ambiguity see also \citet{kie16}.}.

Figure \ref{zoom} allows us to identify other, smaller rotations of the EVPA corresponding to different behaviours of polarization degree and optical flux. In particular, a nearly 60\degr\ change of EVPA was observed in about ten days, between JD=2457231 and 2457241, with $P$ reaching a minimum at the rotation inversion point and the optical flux density remaining in a steadily decreasing phase and achieving a minimum just after $P$.

The MOJAVE Project obtained two VLBA polarization measurements\footnote{http://www.physics.purdue.edu/astro/MOJAVE/}
at 15 GHz in the period we are considering, yielding radio EVPAs of 49\degr\ on 2014 January 11 and 22\degr\ on 2014 December 12.
These values are plotted in Fig.\ \ref{pola} (the latter augmented by $360\degr$) and show that there may be some displacement between the directions of the optical and radio polarization. The position angle of the structure of the radio source estimated from the MOJAVE data is about $-40\degr$, so the radio EVPAs are nearly transverse to the jet axis.

\section{Discussion}
In this paper we focussed on two main issues regarding the spectral and polarimetric behaviour of the HBL PG~1553+113.

In \citet{rai15} we found a disagreement between the SEDs built with {\em Swift} and {\em XMM-Newton} data acquired at a short time distance (about 6 weeks) in 2013. The UVOT spectra showed a pronounced curvature, with a steep UV spectrum whose extrapolation to higher energies did not meet the soft XRT spectrum. 
On the contrary, the optical--UV spectrum obtained from the OM data was flat and the connection with the EPIC spectrum was less puzzling. 
To better clarify the matter, in \citet{rai15} we compared the 2013 data with previous {\em Swift}, IUE and HST-COS observations, confirming the UV spectrum steepness showed by the UVOT data.
We then fitted the various radio-to-X-ray SEDs with an inhomogeneous SSC helical jet model \citep{vil99,rai09}, where different X-ray brightness states and spectral shapes are reproduced by changing only three parameters defining the jet geometry and orientation.
The 2015 SEDs analysed in this paper found the source in an intermediate UV-to-X-ray brightness state with respect to those discussed by \citet{rai15}. They show a consistency within the uncertainties between the {\em XMM-Newton} and {\em Swift} data and smaller optical--UV and X-ray spectral curvatures. 
The two X-ray brightest SEDs are reproduced reasonably well by the simple log-parabolic model that best fits the OM and EPIC data together, while a possibly better fit is still obtained with the helical jet model.
Further {\em XMM-Newton} long exposures on PG~1553+113 are planned in 2017 when, according to the quasi-periodic behaviour claimed by \citet{ack15}, the source will be in outburst. It will be interesting then to study in detail its spectral variability during the high flux state and compare it with the results obtained in this paper relative to a low flux state.

\begin{figure}
   \centering
   \includegraphics[width=84mm]{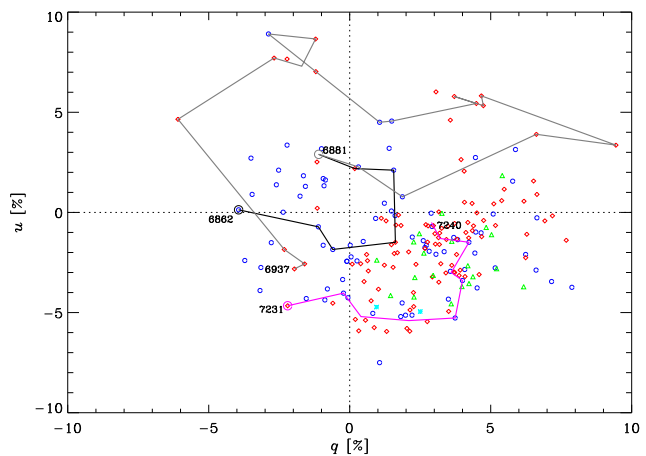}
   \caption{The polarimetric behaviour of PG 1553+113 in the $u$ versus $q$ plot. Red diamonds, cyan asterisks, blue circles and green triangles refer to data from the Crimean, Lulin, Skinakas, and St.~Petersburg observatories, respectively.
Solid lines connect daily averages in chronological order in the time intervals delimited by the epochs indicated in the plot. The counter-clockwise rotation from JD=2456881 to 2457231 (grey) is the continuation of that from JD=2456862 to 2456881 (black).
}
    \label{uvsq}
    \end{figure}

As for the polarimetric behaviour, the major point is the interpretation of the ample EVPA rotation observed in 2014. This is also shown in the $u$ versus $q$ plot in Fig.\ \ref{uvsq}. 
Wide rotations of the radio and optical EVPA have been observed in blazars for more than 30 years \citep[e.g.][]{led79,all81,kik88a,mar08,lar08,mar10,rai12,sas12,rai13,car15,bli15,lar16}. Sometimes they are coincident with $\gamma$-ray flares.
Many models have been proposed to explain EVPA rotations, based on relativistic aberration or on true physical rotations or curved motions in the source emitting region. \citet{bla79} analysed the EVPA swing produced in an accelerating source. A more complex model was proposed by \citet{bjo82}, considering a 3D magnetic field distribution. Rapid changes of the EVPA are predicted to occur close to a minimum in $P$ and in both models the maximum observable amplitude is 180\degr. An interpretation for wider rotations was given by \citet{kon85} in terms of shock waves propagating in a relativistic jet with nonaxysimmetric magnetic field. This can produce in particular a ``steplike" behaviour where minimum $P$ values are expected during the fastest swings (the jumps), as observed in the blazar $0727-115$ by \citet{all81}.
Assuming a shock-in-jet model and a helical magnetic field, \citet{zha14} could explain large EVPA rotations correlated with flux flares, while \citet{zha16} analysed the competition between the shock speed and the magnetization in the emission region and found that in a moderately magnetized environment a fast shock can produce a 180\degr\ rotation with only a modest flux increase. The time scale of the rotation is about a couple of weeks.
EVPA swings from emitting blobs propagating on curved trajectories have been investigated by \citet{nal10}. The maximum rotation rate occurs when the viewing angle is minimum, which implies a minimum in $P$ and a maximum in flux. 
\citet{lar13} satisfactorily fitted the photopolarimetric behaviour of 0716+714 during the 2011 outburst with a model involving the propagation of a shock in a helical jet, which produces a steplike rotation of the EVPA.
Three wide EVPA rotations in different directions were observed by \citet{ale14} when analysing the polarimetric behaviour of PKS $1510-089$ in early 2012. They ascribed this behaviour to turbulence.

The possibility that EVPA rotations are produced by a stochastic process, because of the random walk of the polarization vector due to a turbulent magnetic field, was investigated by e.g.\ \citet{mar14}, \citet{bli15}, and \citet{kie16}. They found that wide rotations can be obtained, but they are rare, so it is unlikely that all those observed are produced by this mechanism. Moreover, stochastic and deterministic processes can both be responsible for different events in the same source.

The extremely wide EVPA rotation of PG 1553+113 in 2014 is similar to that observed in the source $0727-115$ by \citet{all81}, though on a different time scale. There is a general increase, whose finer structure reveals a series of plateaux linked by abrupt jumps. 
The degree of polarization shows a (possibly double) minimum during the rotation that is much wider than 180\degr, and the flux density does not show any flare. 
This questions most of the proposed deterministic models. Hence, it is unlikely that a single emission contribution is responsible for both flux and polarization variability.

\begin{figure}
\includegraphics[width=84mm]{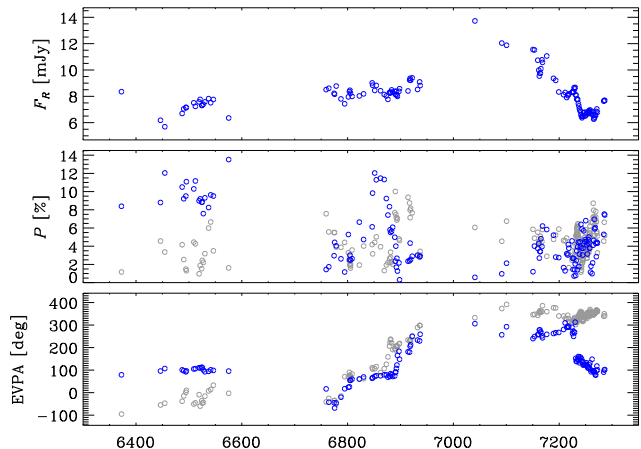}
\caption{The observed $R$-band flux density (top) and the simulated $P$ (middle) and EVPA (bottom) time behaviour (blue circles). They are the results of a Monte Carlo simulation with 220 cells having uniform, but randomly oriented magnetic field. Grey circles show the observed $P$ and EVPA values for a comparison. The 2014 EVPA rotation is well reproduced.}
\label{simu}
\end{figure}

Therefore, following the suggestions by \citet{dar07}, \citet{kie13}, \citet{bli15}, and \citet{kie16}, we assumed that some amount of magnetic turbulence is at work in the optical emitting region and
we ran Monte Carlo simulations to see whether we can reproduce a photopolarimetric behaviour resembling that shown in Fig.\ \ref{pola}. 
We considered the daily binned $R$-band light curve made by photopolarimetric data and assumed that the observed flux is equally contributed by $N_{\rm cells}=220$ emitting cells with randomly oriented magnetic field.
These cells have uniform magnetic field with degree of polarization corresponding to synchrotron radiation from particles with a power-law distribution of energies with index $p$: $P_{\rm max}=(p+1)/(p+7/3)=75\%$ for $p=3$.
At each time step, the flux density is fixed by the observed value, while a number of randomly chosen cells, $N_{\rm var}$, change the orientation of their magnetic field. This number is crucial to obtain a certain level of ordered behaviour.
We set their minimum number to $N_{\rm cells}/10$ and increased them proportionally to the length of the time step $\Delta t$. 
The overall values of $P$ and EVPA at each epoch can be derived from the Stokes parameters $Q$ and $U$ of the whole emitting region, which are obtained by summing the $Q_i$ and $U_i$ parameters of all cells.
We could obtain a variety of different behaviours, with $P$ values statistically close to the observed ones ($P_{\rm min}=1\%, P_{\rm max}=10\%, <P>=4\%, \sigma(P)\sim2 \%$) and sometimes leading to wide EVPA rotations.
We show in Fig.\ \ref{simu} the results of one of these simulations, which produces a wide rotation that is similar to that observed in 2014.
A full statistical analysis is beyond the scope of this paper; we just want to stress here that the polarimetric behaviour shown by PG~1553+113 is difficult to explain in terms of simple emission models while it is easily accounted for by assuming some turbulence of the magnetic field in the emitting region.

Finally, we noticed that the radio EVPAs measured by the MOJAVE project before and after the 2014 optical EVPA rotation are nearly transverse to the jet axis and possibly present some offset with respect to the optical EVPAs.

\section{Summary and conclusions}

We have presented an analysis of the huge multiwavelength observing effort on the blazar PG~1553+113 in 2015, which was motivated by the long pointing time spent by {\em XMM-Newton} during six epochs in July--September.

The EPIC X-ray spectra are well fitted by a simple power-law model with free absorption, yielding neutral hydrogen column densities from 23\% to 52\% higher than the Galactic value.
Fits with a log-parabolic model with Galactic absorption indicate slight spectral curvature.
The long exposures reveal X-ray variability on hourly time scale.
The application of a log-parabolic model with free reddening and absorption to the joint OM+EPIC data suggests a gas/dust ratio that is from 1.5 to 2.6 times higher than in the Milky Way.
During the August 8--10 pointing, simultaneous observations by {\em Swift} were obtained to check inter-calibrations. Both the optical--UV and X-ray data of the two satellites show consistency within the uncertainties. 

The radio-to-optical data acquired by the WEBT collaboration complemented the space information at low energies. The optical and near-infrared light curves show a decreasing trend, in agreement with the UVOT and OM data, while the radio flux displays only modest variability.

We have built the radio-to-X-ray SEDs corresponding to the six {\em XMM-Newton} epochs with simultaneous data from space and ground. We wanted to investigate the properties of the synchrotron emission, and in particular to verify the critical connection between the UV and X-ray spectra that led us to propose an inhomogeneous SSC helical jet model to interpret the SED variability \citep{rai15}. The new SEDs presented in this paper show less curvature in both the optical--UV and X-ray spectra, and an intermediate UV--X-ray flux, as compared to those analysed by \citet{rai15}. As a consequence, at least the two SEDs corresponding to the brightest X-ray states are satisfactorily reproduced by the simple log-parabolic model that best fits the {\em XMM-Newton} data. However,  the helical jet model is still possibly giving a better fit.

We have also analysed the optical polarimetric behaviour of the source over the last three years, along with the X-ray and optical photometric behaviour.
We did not find any clear correlation between the polarization degree and the optical brightness.
The EVPA showed an almost complete counter-clockwise rotation in 2014. In the same period the degree of polarization was very variable and the optical (and possibly the X-ray) flux was rather stable. All together, these pieces of information make an interpretation in terms of simple deterministic emission models quite difficult, while they are more easily explained by a stochastic model that assumes some turbulence of the magnetic field in the jet emitting region.
This is in line with the results of \citet{bli16} and \citet{ang16}, who found that EVPA rotations in high-synchrotron-peaked sources are less frequent than in low-peaked ones and suggested that probably the former are mostly random walk rotations while the latter are due to deterministic processes.

\section*{Acknowledgments}
The data collected by the WEBT collaboration are stored in the WEBT archive at the Osservatorio Astrofisico di Torino - INAF (http://www.oato.inaf.it/blazars/webt/); for questions regarding their availability, please contact the WEBT President Massimo Villata ({\tt villata@oato.inaf.it}).
This research has made use of data obtained through the High Energy Astrophysics Science Archive Research Center Online Service, provided by the NASA/Goddard Space Flight Center.
This research has made use of the XRT Data Analysis Software (XRTDAS) developed under the responsibility of the ASI Science Data Center (ASDC), Italy.
This research was partially funded by the Italian Ministry for Research and Scuola Normale Superiore.
This article is partly based on observations made with the telescopes IAC80 and TCS operated by the Instituto de Astrofisica de Canarias in the Spanish Observatorio del Teide on the island of Tenerife. Most of the observations were taken under the rutinary observation programme. The IAC team acknowledges the support from the group of support astronomers and telescope operators of the Observatorio del Teide.
Based (partly) on data obtained with the STELLA robotic telescopes in Tenerife, an AIP facility jointly operated by AIP and IAC.
St.\ Petersburg University team acknowledges support from Russian RFBR grant 15-02-00949 and St.\ Petersburg University research grant 6.38.335.2015.
AZT-24 observations are made within an agreement between Pulkovo, Rome and Teramo observatories.
G.~Damljanovic and O.~Vince gratefully acknowledge the observing grant support from the Institute of Astronomy and Rozhen National Astronomical Observatory, Bulgaria Academy of Sciences, via bilateral joint research project ``Observations of ICRF radio-sources visible in optical domain" (the head is G.~Damljanovic). This work is a part of the Projects No 176011 (``Dynamics and kinematics of celestial bodies and systems"), No 176004 (``Stellar physics") and No 176021 (``Visible and invisible matter in nearby galaxies: theory and observations") supported by the Ministry of Education, Science and Technological Development of the Republic of Serbia.
The U.\ of Crete/FORTH group acknowledges support by the ``RoboPol” project, which is co-funded by the European Social Fund (ESF) and Greek National Resources, and by the European Comission Seventh Framework Programme (FP7) through grants PCIG10-GA-2011-304001 ``JetPop” and PIRSES-GA-2012-31578 ``EuroCal”.
This research was supported partly by funds of the project RD-08-81 of Shumen University.
The Abastumani team acknowledges financial support by the by Shota Rustaveli NSF under contract FR/577/6-320/13.
This research was partially supported by Scientific Research Fund of the Bulgarian Ministry of Education and Sciences under grant DO 02-137 (BIn-13/09). The Skinakas Observatory is a collaborative project of the University of Crete, the Foundation for Research and Technology -- Hellas, and the Max-Planck-Institut f\"ur Extraterrestrische Physik.
The research at Ulugh Beg Astronomical Institute was funded in Uzbekistan Academy of Sciences grants N.\ F2-FA-F027 and F.4-16.
The Mets\"ahovi team acknowledges the support from the Academy of Finland to our observing projects (numbers 212656, 210338, 121148, and others).
Based on observations with the Medicina telescope operated by INAF-Istituto di Radioastronomia.

\bibliographystyle{mn2e}

\vspace{1cm}\noindent
{\large \bf AFFILIATIONS}

\vspace{0.5cm}\noindent
{\it
$^{ 1}$INAF, Osservatorio Astrofisico di Torino, via Osservatorio 20, I-10025 Pino Torinese, Italy                                                           \\
$^{ 2}$INAF, Osservatorio Astronomico di Roma, via di Frascati 33, I-00040 Monte Porzio Catone, Italy                                                        \\
$^{ 3}$Scuola Normale Superiore, Piazza dei Cavalieri 7, I-56126 Pisa, Italy                                                                                 \\
$^{ 4}$ASI Science Data Center, via del Politecnico s.n.c., I-00133 Roma, Italy                                                                              \\
$^{ 5}$Astronomical Institute, St.\ Petersburg State University, Universitetsky pr. 28, Petrodvoretz, 198504 St.\ Petersburg, Russia                         \\
$^{ 6}$Pulkovo Observatory, 196140 St.\ Petersburg, Russia                                                                                                   \\
$^{ 7}$Department of Physics and ITCP, University of Crete, 71003, Heraklion, Greece                                                                         \\
$^{ 8}$Foundation for Research and Technology - Hellas, IESL, Voutes, 71110 Heraklion, Greece                                                                \\
$^{ 9}$Instituto de Astrofisica de Canarias (IAC), La Laguna, E-38200 Tenerife, Spain                                                                        \\
$^{10}$Departamento de Astrofisica, Universidad de La Laguna, La Laguna, E-38205 Tenerife, Spain                                                             \\
$^{11}$Institute of Astronomy and NAO, Bulgarian Academy of Sciences, 72 Tsarigradsko shosse Blvd., 1784 Sofia, Bulgaria                                     \\
$^{12}$Crimean Astrophysical Observatory RAS, P/O Nauchny, 298409, Russia                                                                                    \\
$^{13}$EPT Observatories, Tijarafe, E-38780 La Palma, Spain                                                                                                  \\
$^{14}$INAF, TNG Fundaci\'on Galileo Galilei, E-38712 La Palma, Spain                                                                                        \\
$^{15}$Graduate Institute of Astronomy, National Central University, 300 Jhongda Rd, Jhongli City, Taoyuan County 32001, Taiwan (R.O.C.)                     \\
$^{16}$Astronomical Observatory, Volgina 7, 11060 Belgrade, Serbia                                                                                           \\
$^{17}$Ulugh Beg Astronomical Institute, Maidanak Observatory, Uzbekistan                                                                                    \\
$^{18}$INAF, Osservatorio Astrofisico di Catania, Via S.\ Sofia 78, I-95123 Catania, Italy                                                                   \\
$^{19}$INAF, Istituto di Radioastronomia, via Gobetti 101, I-40129 Bologna, Italy                                                                            \\
$^{20}$Department of Theoretical and Applied Physics, University of Shumen, 9712 Shumen, Bulgaria                                                            \\
$^{21}$Abastumani Observatory, Mt. Kanobili, 0301 Abastumani, Georgia                                                                                        \\
$^{22}$Engelhardt Astronomical Observatory, Kazan Federal University, Tatarstan, Russia                                                                      \\
$^{23}$Aalto University Mets\"ahovi Radio Observatory, Mets\"ahovintie 114, 02540 Kylm\"al\"a, Finland                                                       \\
$^{24}$Aalto University Department of Radio Science and Engineering, P.O. BOX 13000, FI-00076 AALTO, Finland                                                 \\
$^{25}$Department of Physics and Astronomy, Brigham Young University, Provo, UT 84602, USA                                                                   \\
$^{26}$Michael Adrian Observatorium, Astronomie Stiftung Trebur, 65468 Trebur, Germany                                                                       \\
$^{27}$University of Applied Sciences, Technische Hochschule Mittelhessen, 61169 Friedberg, Germany                                                          \\
$^{28}$Osservatorio Astronomico Sirio, Piazzale Anelli, I-70013 Castellana Grotte, Italy                                                                     \\
$^{29}$Department of Physics, University of Colorado Denver, CO, 80217-3364 USA                                                                              \\
}
\bsp

\label{lastpage}

\end{document}